
%
%
\def\unredoffs{} \def\redoffs{\voffset=-.31truein\hoffset=-.59truein}
\def\speclscape{}
%
%
%
%
\newbox\leftpage \newdimen\fullhsize \newdimen\hstitle \newdimen\hsbody
\tolerance=1000\hfuzz=2pt
\catcode`\@=11 
\def\bigans{b }
\def\answ{b }
%

\ifx\answ\bigans\message{(This will come out unreduced.}
\magnification=1200\unredoffs\baselineskip=16pt plus 2pt minus 1pt
\hsbody=\hsize \hstitle=\hsize 
\else\message{(This will be reduced.} \let\l@r=L
\magnification=1000\baselineskip=16pt plus 2pt minus 1pt \vsize=7truein
\redoffs \hstitle=8truein\hsbody=4.75truein\fullhsize=10truein\hsize=\hsbody
\output={\ifnum\pageno=0 
  \shipout\vbox{\speclscape{\hsize\fullhsize\makeheadline}
   \hbox to \fullhsize{\hfill\pagebody\hfill}}\advancepageno
  \else
 \almostshipout{\leftline{\vbox{\pagebody\makefootline}}}\advancepageno
  \fi}
\def\almostshipout#1{\if L\l@r \count1=1 \message{[\the\count0.\the\count1]}
      \global\setbox\leftpage=#1 \global\let\l@r=R
 \else \count1=2
  \shipout\vbox{\speclscape{\hsize\fullhsize\makeheadline}
      \hbox to\fullhsize{\box\leftpage\hfil#1}}  \global\let\l@r=L\fi}
\fi
%
\newcount\yearltd\yearltd=\year\advance\yearltd by -1900

\def\Title#1#2{\nopagenumbers\abstractfont\hsize=\hstitle\rightline{#1}%
\vskip 1in\centerline{\titlefont #2}\abstractfont\vskip .5in\pageno=0}
\def\Date#1{\vfill\leftline{#1}\tenpoint\supereject\global\hsize=\hsbody%
\footline={\hss\tenrm\folio\hss}}
%

\def\draftmode{\message{ DRAFTMODE }\def\draftdate{{\rm preliminary draft:
\number\month/\number\day/\number\yearltd\ \ \hourmin}}%
\headline={\hfil\draftdate}\writelabels\baselineskip=20pt plus 2pt minus 2pt
 {\count255=\time\divide\count255 by 60 \xdef\hourmin{\number\count255}
  \multiply\count255 by-60\advance\count255 by\time
  \xdef\hourmin{\hourmin:\ifnum\count255<10 0\fi\the\count255}}}
\def\nolabels{\def\wrlabeL##1{}\def\eqlabeL##1{}\def\reflabeL##1{}}
\def\writelabels{\def\wrlabeL##1{\leavevmode\vadjust{\rlap{\smash%
{\line{{\escapechar=` \hfill\rlap{\sevenrm\hskip.03in\string##1}}}}}}}%
\def\eqlabeL##1{{\escapechar-1\rlap{\sevenrm\hskip.05in\string##1}}}%
\def\reflabeL##1{\noexpand\llap{\noexpand\sevenrm\string\string\string##1}}}
\nolabels
%
\global\newcount\secno \global\secno=0
\global\newcount\meqno \global\meqno=1
\def\newsec#1{\global\advance\secno by1\message{(\the\secno. #1)}
\global\subsecno=0\eqnres@t\noindent{\bf\the\secno. #1}
\writetoca{{\secsym} {#1}}\par\nobreak\medskip\nobreak}
\def\eqnres@t{\xdef\secsym{\the\secno.}\global\meqno=1\bigbreak\bigskip}
\def\sequentialequations{\def\eqnres@t{\bigbreak}}\xdef\secsym{}
\global\newcount\subsecno \global\subsecno=0
\def\subsec#1{\global\advance\subsecno by1\message{(\secsym\the\subsecno. #1)}
\ifnum\lastpenalty>9000\else\bigbreak\fi
\noindent{\it\secsym\the\subsecno. #1}\writetoca{\string\quad
{\secsym\the\subsecno.} {#1}}\par\nobreak\medskip\nobreak}
\def\appendix#1#2{\global\meqno=1\global\subsecno=0\xdef\secsym{\hbox{#1.}}
\bigbreak\bigskip\noindent{\bf Appendix #1. #2}\message{(#1. #2)}
\writetoca{Appendix {#1.} {#2}}\par\nobreak\medskip\nobreak}
%
%
\def\eqnn#1{\xdef #1{(\secsym\the\meqno)}\writedef{#1\leftbracket#1}%
\global\advance\meqno by1\wrlabeL#1}
\def\eqna#1{\xdef #1##1{\hbox{$(\secsym\the\meqno##1)$}}
\writedef{#1\numbersign1\leftbracket#1{\numbersign1}}%
\global\advance\meqno by1\wrlabeL{#1$\{\}$}}
\def\eqn#1#2{\xdef #1{(\secsym\the\meqno)}\writedef{#1\leftbracket#1}%
\global\advance\meqno by1$$#2\eqno#1\eqlabeL#1$$}
%
\newskip\footskip\footskip14pt plus 1pt minus 1pt 
\def\footnotefont{\ninepoint}\def\f@t#1{\footnotefont #1\@foot}
\def\f@@t{\baselineskip\footskip\bgroup\footnotefont\aftergroup\@foot\let\next}
\setbox\strutbox=\hbox{\vrule height9.5pt depth4.5pt width0pt}
\global\newcount\ftno \global\ftno=0
\def\foot{\global\advance\ftno by1\footnote{$^{\the\ftno}$}}
%
\newwrite\ftfile
\def\footend{\def\foot{\global\advance\ftno by1\chardef\wfile=\ftfile
$^{\the\ftno}$\ifnum\ftno=1\immediate\openout\ftfile=foots.tmp\fi%
\immediate\write\ftfile{\noexpand\smallskip%
\noexpand\item{f\the\ftno:\ }\pctsign}\findarg}%
\def\footatend{\vfill\eject\immediate\closeout\ftfile{\parindent=20pt
\centerline{\bf Footnotes}\nobreak\bigskip\input foots.tmp }}}
\def\footatend{}
%
%
\global\newcount\refno \global\refno=1
\newwrite\rfile
\def\ref{[\the\refno]\nref}
\def\nref#1{\xdef#1{[\the\refno]}\writedef{#1\leftbracket#1}%
\ifnum\refno=1\immediate\openout\rfile=refs.tmp\fi
\global\advance\refno by1\chardef\wfile=\rfile\immediate
\write\rfile{\noexpand\item{#1\ }\reflabeL{#1\hskip.31in}\pctsign}\findarg}
\def\findarg#1#{\begingroup\obeylines\newlinechar=`\^^M\pass@rg}
{\obeylines\gdef\pass@rg#1{\writ@line\relax #1^^M\hbox{}^^M}%
\gdef\writ@line#1^^M{\expandafter\toks0\expandafter{\striprel@x #1}%
\edef\next{\the\toks0}\ifx\next\em@rk\let\next=\endgroup\else\ifx\next\empty%
\else\immediate\write\wfile{\the\toks0}\fi\let\next=\writ@line\fi\next\relax}}
\def\striprel@x#1{} \def\em@rk{\hbox{}}
\def\lref{\begingroup\obeylines\lr@f}
\def\lr@f#1#2{\gdef#1{\ref#1{#2}}\endgroup\unskip}

\def\addref#1{\immediate\write\rfile{\noexpand\item{}#1}} 
\def\startrefs#1{\immediate\openout\rfile=refs.tmp\refno=#1}
\def\xref{\expandafter\xr@f}\def\xr@f[#1]{#1}
\def\refs#1{\count255=1[\r@fs #1{\hbox{}}]}
\def\r@fs#1{\ifx\und@fined#1\message{reflabel \string#1 is undefined.}%
\nref#1{need to supply reference \string#1.}\fi%
\vphantom{\hphantom{#1}}\edef\next{#1}\ifx\next\em@rk\def\next{}%
\else\ifx\next#1\ifodd\count255\relax\xref#1\count255=0\fi%
\else#1\count255=1\fi\let\next=\r@fs\fi\next}
%

%
\newwrite\ffile\global\newcount\figno \global\figno=1
\def\fig{Figure~\the\figno\nfig}
\def\nfig#1{\xdef#1{Figure~\the\figno}%
\writedef{#1\leftbracket fig.\noexpand~\the\figno}%
\ifnum\figno=1\immediate\openout\ffile=figs.tmp\fi\chardef\wfile=\ffile%
\immediate\write\ffile{\noexpand\medskip\noexpand\item{Fig.\ \the\figno. }
\reflabeL{#1\hskip.55in}\pctsign}\global\advance\figno by1\findarg}
\def\xfig{\expandafter\xf@g}\def\xf@g fig.\penalty\@M\ {}
\def\figs#1{figs.~\f@gs #1{\hbox{}}}
\def\f@gs#1{\edef\next{#1}\ifx\next\em@rk\def\next{}\else
\ifx\next#1\xfig #1\else#1\fi\let\next=\f@gs\fi\next}
\newwrite\lfile
{\escapechar-1\xdef\pctsign{\string\%}\xdef\leftbracket{\string\{}
\xdef\rightbracket{\string\}}\xdef\numbersign{\string\#}}

\def\writestop{\def\writestoppt{\immediate\write\lfile{\string\pageno%
\the\pageno\string\startrefs\leftbracket\the\refno\rightbracket%
\string\def\string\secsym\leftbracket\secsym\rightbracket%
\string\secno\the\secno\string\meqno\the\meqno}\immediate\closeout\lfile}}
\def\writestoppt{}\def\writedef#1{}
\def\seclab#1{\xdef #1{\the\secno}\writedef{#1\leftbracket#1}\wrlabeL{#1=#1}}
\def\subseclab#1{\xdef #1{\secsym\the\subsecno}%
\writedef{#1\leftbracket#1}\wrlabeL{#1=#1}}
\newwrite\tfile \def\writetoca#1{}
\def\leaderfill{\leaders\hbox to 1em{\hss.\hss}\hfill}
\def\writetoc{\immediate\openout\tfile=toc.tmp
   \def\writetoca##1{{\edef\next{\write\tfile{\noindent ##1
   \string\leaderfill {\noexpand\number\pageno} \par}}\next}}}

%
\catcode`\@=12 
%
\edef\tfontsize{\ifx\answ\bigans scaled\magstep3\else scaled\magstep4\fi}
\font\titlerm=cmr10 \tfontsize \font\titlerms=cmr7 \tfontsize
\font\titlermss=cmr5 \tfontsize \font\titlei=cmmi10 \tfontsize
\font\titleis=cmmi7 \tfontsize \font\titleiss=cmmi5 \tfontsize
\font\titlesy=cmsy10 \tfontsize \font\titlesys=cmsy7 \tfontsize
\font\titlesyss=cmsy5 \tfontsize \font\titleit=cmti10 \tfontsize
\skewchar\titlei='177 \skewchar\titleis='177 \skewchar\titleiss='177
\skewchar\titlesy='60 \skewchar\titlesys='60 \skewchar\titlesyss='60
\def\titlefont{\def\rm{\fam0\titlerm}
\textfont0=\titlerm \scriptfont0=\titlerms \scriptscriptfont0=\titlermss
\textfont1=\titlei \scriptfont1=\titleis \scriptscriptfont1=\titleiss
\textfont2=\titlesy \scriptfont2=\titlesys \scriptscriptfont2=\titlesyss
\textfont\itfam=\titleit \def\it{\fam\itfam\titleit}\rm}
 \ifx\answ\bigans\else scaled\magstep1\fi
\ifx\answ\bigans\def\abstractfont{\tenpoint}\else
\font\abssl=cmsl10 scaled \magstep1
\font\absrm=cmr10 scaled\magstep1 \font\absrms=cmr7 scaled\magstep1
\font\absrmss=cmr5 scaled\magstep1 \font\absi=cmmi10 scaled\magstep1
\font\absis=cmmi7 scaled\magstep1 \font\absiss=cmmi5 scaled\magstep1
\font\abssy=cmsy10 scaled\magstep1 \font\abssys=cmsy7 scaled\magstep1
\font\abssyss=cmsy5 scaled\magstep1 \font\absbf=cmbx10 scaled\magstep1
\skewchar\absi='177 \skewchar\absis='177 \skewchar\absiss='177
\skewchar\abssy='60 \skewchar\abssys='60 \skewchar\abssyss='60
\def\abstractfont{\def\rm{\fam0\absrm}
\textfont0=\absrm \scriptfont0=\absrms \scriptscriptfont0=\absrmss
\textfont1=\absi \scriptfont1=\absis \scriptscriptfont1=\absiss
\textfont2=\abssy \scriptfont2=\abssys \scriptscriptfont2=\abssyss
\textfont\itfam=\bigit \def\it{\fam\itfam\bigit}\def\footnotefont{\tenpoint}%
\textfont\slfam=\abssl \def\sl{\fam\slfam\abssl}%
\textfont\bffam=\absbf \def\bf{\fam\bffam\absbf}\rm}\fi
\def\tenpoint{\def\rm{\fam0\tenrm}
\textfont0=\tenrm \scriptfont0=\sevenrm \scriptscriptfont0=\fiverm
\textfont1=\teni  \scriptfont1=\seveni  \scriptscriptfont1=\fivei
\textfont2=\tensy \scriptfont2=\sevensy \scriptscriptfont2=\fivesy
\textfont\itfam=\tenit \def\it{\fam\itfam\tenit}\def\footnotefont{\ninepoint}%
\textfont\bffam=\tenbf \def\bf{\fam\bffam\tenbf}\def\sl{\fam\slfam\tensl}\rm}
\font\ninerm=cmr9 \font\sixrm=cmr6 \font\ninei=cmmi9 \font\sixi=cmmi6
\font\ninesy=cmsy9 \font\sixsy=cmsy6 \font\ninebf=cmbx9
\font\nineit=cmti9 \font\ninesl=cmsl9 \skewchar\ninei='177
\skewchar\sixi='177 \skewchar\ninesy='60 \skewchar\sixsy='60
\def\ninepoint{\def\rm{\fam0\ninerm}
\textfont0=\ninerm \scriptfont0=\sixrm \scriptscriptfont0=\fiverm
\textfont1=\ninei \scriptfont1=\sixi \scriptscriptfont1=\fivei
\textfont2=\ninesy \scriptfont2=\sixsy \scriptscriptfont2=\fivesy
\textfont\itfam=\ninei \def\it{\fam\itfam\nineit}\def\sl{\fam\slfam\ninesl}%
\textfont\bffam=\ninebf \def\bf{\fam\bffam\ninebf}\rm}
%
%

\hyphenation{anom-aly anom-alies coun-ter-term coun-ter-terms}
\def\inv{^{\raise.15ex\hbox{${\scriptscriptstyle -}$}\kern-.05em 1}}

\def\Dsl{\,\raise.15ex\hbox{/}\mkern-13.5mu D} 
\def\dsl{\raise.15ex\hbox{/}\kern-.57em\partial}

\font\bigit=cmti10 scaled \magstep1
\def\lspace{\ifx\answ\bigans{}\else\qquad\fi}
\def\lbspace{\ifx\answ\bigans{}\else\hskip-.2in\fi} 
\def\boxeqn#1{\vcenter{\vbox{\hrule\hbox{\vrule\kern3pt\vbox{\kern3pt
	\hbox{${\displaystyle #1}$}\kern3pt}\kern3pt\vrule}\hrule}}}
\def\mbox#1#2{\vcenter{\hrule \hbox{\vrule height#2in
		\kern#1in \vrule} \hrule}}  
%

\def\darr#1{\raise1.5ex\hbox{$\leftrightarrow$}\mkern-16.5mu #1}

\def\half{{\textstyle{1\over2}}} 
\def\roughly#1{\raise.3ex\hbox{$#1$\kern-.75em\lower1ex\hbox{$\sim$}}}

\def\curl{\nabla\times}

\def\dot{\cdot}
\def\perpp{{\!\scriptscriptstyle\perp}}

\def\half{{1\over 2}}

\def\dj{\partial_j}

\def\dk{\partial_k}
\def\dn{\delta n}
\def\dx{\partial_x}
\def\dy{\partial_y}
\def\dz{\partial_z}
\def\angstrom{\AA}

\def\bo#1{{\cal O}(#1)}
\def\cross{\times}

\def\dnb{\delta\vec n}
\def\nabb{\nabla}

\def\free{\hbox{$\cal F$}}
\def\bold#1{\setbox0=\hbox{$#1$}%
     \kern-.010em\copy0\kern-\wd0
     \kern.025em\copy0\kern-\wd0
     \kern-.020em\raise.0200em\box0 }
\def\ts{\theta_6}

\lref\KLEMi{M. Kl\'eman, J. Phys. (Paris) {\bf 46} (1985) 1193.}
\lref\SWM{S.~Meiboom, J.P.~Sethna, P.W.~Anderson and W.F.~Brinkman, Phys. Rev.
Lett. {\bf 46}
(1981) 1216; see also
D.C.~Wright and N.D.~Mermin, Rev. Mod. Phys. {\bf 61} (1989) 385.}
\lref\KLEM{M.~Kl\'eman, Rep. Prog. Phys. {\bf 52} (1989) 555.}
\lref\NT{D.R.~Nelson and J.~Toner, Phys. Rev. B {\bf 24} (1981) 363.}
\lref\MH{N.D.~Mermin and T.L.~Ho, Phys. Rev. Lett. {\bf 36} (1976) 594.}
\lref\TER{E.M.~Terentjev, Europhys. Lett. {\bf 23} (1993) 27.}
\lref\GIA{C.~Gianessi, Phys. Rev. A {\bf 28} (1983) 350; Phys. Rev. A {\bf
34} (1986) 705.}
\lref\MN{M.C.~Marchetti and D.R.~Nelson, Phys. Rev. B {\bf 41} (1990) 1910.}
\lref\VOL{G.E.~Volovik, {\sl Exotic Properties of $^3$He}, Chap. III
(World Scientific, Singapore, 1992).}

\lref\TON{J.~Toner, Phys. Rev. A {\bf 27} (1983) 1157.}
\lref\HN{B.I.~Halperin and D.R.~Nelson, Phys. Rev. Lett. {\bf 41} (1978) 121;
D.R.~Nelson
and B.I.~Halperin, Phys. Rev. B {\bf 19} (1979) 2457.}
\lref\SL{S.~Langer and J.~Sethna, Phys. Rev. A {\bf 34}
(1986) 5035; G.A.~Hinshaw, Jr., R.G.~Petschek and R.A.~Pelcovits, Phys. Rev.
Lett.
{\bf 60}
(1988) 1864.}
\lref\PN{P.~Nelson and T.~Powers, Phys. Rev. Lett. {\bf 69} (1992) 3409;
J. Phys. II (Paris) {\bf 3} (1993) 1535.}
\lref\SEL{J.V.~Selinger and J.M.~Schnur, Phys. Rev. Lett. {\bf 71}
(1993) 4091; J.V. Selinger, Z.-G. Wang, R.F.~Bruinsma and C.M.~Knobler,
Phys. Rev. Lett. {\bf 70} (1993) 1139.}
\lref\KN{R.D.~Kamien and D.R.~Nelson, Phys. Rev. Lett. {\bf 74} (1995) 2499;
Institute for Advanced Study Preprint
IASSNS-HEP-95/10, in preparation (1995).}
\lref\TGB{S.R.~Renn and T.C.~Lubensky, Phys. Rev. A {\bf 38} (1988) 2132;
T.C.~Lubensky
and S.R.~Renn, Phys. Rev. A
{\bf 41} (1990) 4392.}
\lref\MEYER{R.B.~Meyer, Appl. Phys. Lett. {\bf 12}, 281 (1968); Appl. Phys.
Lett. {\bf 14}
(1969) 208.}
\lref\TONER{J.~Toner, Phys. Rev. Lett. {\bf 68} (1992) 1331.}
\lref\KT{R.D.~Kamien and J.~Toner, Phys. Rev. Lett. {\bf 74} (1995) 3181.}
\lref\NELii{D.R.~Nelson and L.~Peliti,  J. Phys. (Paris) {\bf 48} (1987) 1085;
see also F.~David, {\sl Statistical Mechanics of Membranes and Surfaces :
Jerusalem Winter School for Theoretical Physics}, edited by D.R.~Nelson,
{\sl et. al.} (World Scientific, Singapore, 1989).}

\Title{IASSNS-HEP-95/44}{Liquids with Chiral Bond Order}

\centerline{Randall D. Kamien\footnote{$^\dagger$}
{\baselineskip .18truein Address after 1 August 1995: Department of Physics
and Astronomy, University of Pennsylvania, Philadelphia, PA 19104\hfill\break
email: \tt kamien@lubensky.physics.upenn.edu}}
\bigskip\centerline{\sl School of Natural Sciences,
Institute for Advanced Study, Princeton, NJ 08540}

\vskip .3in
I describe new phases of a chiral liquid crystal with nematic and hexatic
order.  I find a conical phase, similar to that of a cholesteric in an applied
magnetic field for Frank elastic constants $K_2>K_3$.  I discuss the role
of fluctuations in the context of this phase and the possibility
of satisfying the inequality for sufficiently long polymers.
In addition I discuss the topological constraint relating defects
in the bond order field to textures of the nematic and elucidate its physical
meaning.  Finally I discuss the analogy between smectic liquid crystals
and chiral hexatics and propose
a defect-riddled ground state, akin to the Renn-Lubensky
twist grain boundary phase of chiral smectics.

\Date{19 July 1995}
\newsec{Introduction and Summary}

Since time immemorial \TON\ people have looked for
liquid crystal phases with bond-orientational
order \HN .  Aside from the possibility of a new type of liquid crystal,
hexatic-type
order can be an intermediate stage in the {\sl continuous} freezing of a liquid
in
three dimensions.  Much more recently there has been an explosive progress in
new
chiral phases in both two \refs{\SL,\SEL,\PN}, and three dimensions
\refs{\TGB,\KN}.
In this paper I propose both a uniform and defect laden ground state of
a liquid crystal with {\sl both} hexatic order and chirality.

Toner \TON\ has proposed that nematic liquid crystals, upon cooling, could form
a liquid crystalline phase with nematic order and hexatic order in the plane
perpendicular to it.
In section 2 I will consider the melting of a chiral columnar phase \KN\
into a chiral liquid with hexatic order.  I discuss both the Landau theory
as well as the melting of the crystal through dislocation loop unbinding.  When
a proliferation of edge and screw dislocations develops, the crystal melts,
leaving
a normal liquid with hexatic order and no residual constraints
associated with the crystal rigidity.  Typically, long thin molecules do not
have
columnar phases, but freeze via smectic phases.  I expect then that chiral,
disc shaped
molecules could participate in the structures discussed here.

It is well known \MEYER\ that cholesteric phases, under applied magnetic fields
can unwind as well as tip up and become conical phases.  However, the conical
phases exist only when $K_2>K_3$, which is seldom the case experimentally.
When
$K_2<K_3$ there is a discontinuous transition from a cholesteric with
director in the $xy$-plane to a nematic aligned along the field in the $\hat z$
direction.
In section 3 I find that hexatic order has the same effect as the magnetic
field, although
it does not participate in unwinding the helical pitch.  Adapting
the known consequences of fluctuations and
nonlinearities of polymer nematics \TONER\ and cholesterics \KT\  to the chiral
polymer hexatic, I find anomalously large elastic constants which depend on the
polymer length.  For sufficiently long polymers $K_2$ will eventually grow
larger than $K_3$.  If liquid crystal phases can still exist for these long
polymers, Meyer's conical phase would appear.  This would be evidence
of hexatic order.  When the nematic director has
a conical texture, as in a smectic-$C^*$ texture, and
the molecules are chiral, an electric polarization can develop.  Thus
this phase could exhibit ferroelectric liquid crystalline
behavior \ref\MEFE{R.B.~Meyer, L.~Liebert. L.~Strzelecki, and P.~Keller, J.
Phys. (Paris)
Lett {\bf 36} (1975) L-69.} .

Since the bond order is defined in the plane perpendicular to the nematic
director
$\bf\hat n$, its value must be determined in a way which takes into account
the texture of the director.
Because there is a nematic director ${\bf\hat n}$ and a bond-angle
order parameter $\ts$, the non-chiral theory is similar to that of
the $A$ phase of $^3He$.  In section 4 I will discuss
the geometric constraints relating the nematic director
to the bond-order field in the absence of free disclinations.  This is known as
the Mermin-Ho relation in $^3He$ \MH.  It plays an important role in
understanding
the allowed ground states of the chiral N+6 phase and elucidates the structure
of the allowed defects.

With chirality added the N+6 order can exhibit a
large variety of phases similar in structure to known phases of smectics.
When chiral molecules participate in N+6 order, two kinds of twisting are
allowed \KN : twisting of the nematic order, leading to cholesteric states, and
twisting of the bond order, leading to braided moir\'e states.  The twisting
bond order, with an ever increasing angle, is analogous to the layered
order of smectics with the pitch of the bond order equivalent to the layer
spacing
of the smectic.  This allows me to consider defect phases similar to the
Renn-Lubensky
TGB state in chiral smectics.  I discuss this in section 5.

\newsec{Landau Theory and Melting of the Hexagonal Columnar Phase}
I first derive the free energy of the chiral N+6 phase as the sum of its
constituent parts.
Fluctuations in the nematic director are described by the Frank free energy
density:
\eqn\ei{\free_{{\bf\hat n}} = {K_1\over 2}\left(\nabla\cdot{\bf\hat n}\right)^2
+ {K_2\over
2}\left[{\bf\hat n}\cdot(\nabla\!\times\!{\bf\hat n}) - q_0\right]^2 +
{K_3\over
2}\left[{\bf\hat n}\!\times\!(\nabla\!\times\!{\bf\hat n})\right]^2}
where $K_i$ are the Frank elastic constants and $2\pi/q_0$ is the equilibrium
cholesteric
pitch.  In order to define bond-angle order, I must, everywhere in space,
define
a right-handed, orthonormal triad $\left\{{\bf\hat e}_1,{\bf\hat e}_2,
{{\bf\hat n}}\right\}$, where ${\bf\hat e}_2 = {\bf\hat n}\cross{\bf\hat e}_1$.
 In this case,
the usual
hexatic order parameter, $\psi_6 = \vert\psi_6\vert e^{6i\ts}$ is defined
by
\eqn\tsdef{\psi_6({\bf r}) = \sum_{i\in P({\bf\hat n},{\bf r})}
e^{6i\theta_i({\bf r})}}
where $\theta_i({\bf r})$ is the angle between the particle $i$ and the basis
vector ${\bf\hat e}_1$
as measured around $\bf r$ with the sign of the
angle determined by $\bf\hat n$, and $P({\bf\hat n},{\bf r})$ is the set of
particles that are in the plane perpendicular to $\bf\hat n$ through the point
$\bf r$.
The vector ${\bf\hat e}_1$ is chosen unambiguously throughout space (possibly
modulo rotations by $2\pi/6$) and may be thought of as pointing to one of the
ground state
nearest neighbors.  More precisely,
\eqn\eangle{\sin\left[\theta_i({\bf r})\right] = {
{\bf\hat n}\dot\left[{\bf\hat e}_1\cross({\bf r_i -r})\right]\over
\vert\vert{\bf r_i -r}\vert\vert}.}
and the sum \tsdef\ is taken only over particles at position ${\bf r}_i$ such
that
${\bf\hat n}({\bf r})\dot({\bf r}_i -{\bf r})=0$.  The bond-angle order
parameter
now depends on the texture of $\bf\hat n$.  As a result, one must take into
account the nematic when taking derivatives of $\ts$: a covariant derivative
must be employed.  For now I will ignore this complication and will return to
it
in Section 4 and will justify the na\"\i ve analysis {\sl a posteriori}.

Because the definition of
$\ts$ depends on the direction of $\bf\hat n$, under the transformation
${\bf\hat n}\rightarrow -{\bf\hat n}$, \eangle\ implies that $\theta_i
\rightarrow -\theta_i$ and hence $\ts\rightarrow -\ts$
(or equivalently, $\psi_6\leftrightarrow\psi_6^*$).  When constructing
a free energy that includes $\ts$, the overall nematic symmetry ${\bf\hat n}
\rightarrow -{\bf\hat n}$ must be preserved.  Owing to the definition of
$\ts$, the free energy can include any term with even powers of
$\bf\hat n$ and $\ts$ {\sl together}.  The spin wave theory for $\ts$
includes spin-stiffnesses as well as a new chiral term:
\eqn\eii{\free_{\ts} = {K_A^{||}-K_A^\perp\over 2}\left({\bf\hat
n}\cdot\nabla\ts\right)^2
+ {K_A^\perp\over 2}\left(\nabla\ts\right)^2 - K_A^{||}\tilde q_0{\bf\hat
n}\cdot\nabla\ts}
where, since there is a preferred direction ${\bf\hat n}$, I have included the
possibility
of anisotropic stiffnesses.  The final term is chiral and indicates the
tendency
for the bond-angle order parameter to rotate around the nematic director with
pitch $2\pi/\tilde q_0$ \refs{\KN,\TER}.  Finally there
are additional non-chiral
couplings between ${\bf\hat n}$ and $\ts$ \refs{\TON,\GIA}:
\eqn\eiii{\free_{{\bf\hat n}\ts}= \bar C\left({\bf\hat
n}\cdot\nabla\ts\right)\left[{\bf\hat n}\cdot(\nabla
\!\times\!{\bf\hat n})\right] +\bar C'\nabla\ts\cdot\nabla\!\times\!{\bf\hat
n}}
The free energy is the sum of terms $F = \int d^3\!x\,\{\free_{{\bf\hat n}} +
\free_{\ts}
+\free_{{\bf\hat n}\ts}\}$.

Recently the theory of chiral polymer crystals has been formulated \KN .
One would imagine
that the total free energy resulted from the melting of such a crystal.
Indeed, this is the
case.  Considering only quadratic fluctuations around a ground state with
$\langle\hat
n\rangle =\hat z$, the additional free energy coming from the two-dimensional
crystal displacement field $\vec u$ is
\eqn\evi{\free_{\rm crystal} = \mu(\partial_iu_j -\epsilon_{ij}\ts)^2
+ {\lambda\over 2}u_{ii}^2 + \mu'(\partial_z u_i-\delta n_i)^2}
At long wavelengths $\ts$ and $\delta n_i$ are locked into crystal deformations
by
$\ts=\half\epsilon_{ij}\partial_iu_j$ and $\delta n_i=\partial_zu_i$.  Thus
$2\partial_z\ts - \epsilon_{ij}\partial_i\delta n_j =0$.  When dislocations
are introduced into the crystal, derivatives do not commute so
$\partial_z\partial_i\ne\partial_i\partial_z$.  In this case I introduce
$w_{\gamma i}$ which is equal to $\partial_\gamma u_i$ away from dislocations.
As in \MN\ I introduce a dislocation density tensor $\alpha_{\mu i}$,
the density of dislocations running in the $\mu$ direction
with Burgers vector in the $i$ direction and
$\epsilon_{\mu\nu\gamma}\partial_\nu w_{\gamma i}
= -\alpha_{\mu i}$.  With $\delta n_i=w_{zi}$ and
$\ts=\half\epsilon_{ij}w_{ij}$ the
long wavelength constraint between $\ts$ and $\delta n_i$
becomes instead \KN
\eqn\ev{2\partial_z\ts - \epsilon_{ij}\partial_i\delta n_j = - {\rm
Tr}[\alpha]}
where $\epsilon_{ij}$ is the two-dimensional anti-symmetric tensor.
In particular $\ev$ constrains the two
chiral terms to differ only by the density of screw dislocations.

Since
a proliferation of {\sl edge} dislocations running along the $\hat z$ direction
would be sufficient to melt a two-dimensional columnar crystal, one might think
that
the constraint would not be altered and that the free energy of the liquid
state
would be augmented by it.  However, this is not the case.  In the usual
scenario
of two-dimensional melting \HN\ dislocation pairs unbind, leading to a hexatic
liquid.
In
the case of a columnar crystal, parallel dislocation {\sl lines} can unbind,
leading
to a hexatic.  In a columnar crystal, defects
running parallel to the columnar direction ($\hat z$) are edge dislocations.
Those defects
lying in the perpendicular plane ($xy$) are either edge or screw dislocations,
depending
on whether the Burgers vector is normal or parallel to the dislocation line,
respectively.  Thus dislocation loops which contribute to the melting of the
crystal
through edge dislocations along $\hat z$ will contain, as well, screw-like
and edge-like dislocations in the plane.
Since the energy per unit length of a pair of edge dislocations parallel
to $\hat z$
is finite, infinite dislocations,
ending at the boundaries, will not proliferate.
Instead, dislocation {\sl loops} will
unbind \MN\
leading to a proliferation of {\sl screw} dislocations.  To incorporate the
presence of screw dislocations I add to \evi\
a term representing the dislocation free energy density \ref\TONSM{I
thank J.~Toner for discussions on this point.}:
\eqn\eaddtoevi{
\delta\free = E_{\mu i\nu j}({\bf q})\alpha_{\mu i}({\bf q})\alpha_{\nu j}({\bf
q})}
where $E_{\mu i\nu j}$ represents the core and elastic energy of the defects.
In the pure crystal the energy per unit length
of an unpaired edge dislocation parallel to the $\hat z$-axis
diverges logarithmically with the system size.
Dislocation {\sl loops}, however will have a finite energy, scaling like $R\ln
R$ where
$R$ is the size of the loop.  Below the melting transition, where the defect
loops are
bound, the free energy will favor $\alpha_{\mu i}=0$ and
hence ${\rm Tr}[\alpha]=0$ everywhere.  Thus $2\dz\ts = \curl\dnb$.  However,
when dislocation loops unbind, the effective quadratic energy for $\alpha_{\mu
i}$
will be finite.  The diagonal part of \eaddtoevi\ becomes
\eqn\ediagn{\delta\free_{\rm diag} = E_{\rm screw}\left\{{\rm
Tr}[\alpha]\right\}^2 =
E_{\rm screw} \left[2\dz\ts - \curl\dnb\right]^2}
Thus the constraint \ev\ will be demoted to
merely a preference for configurations with
$2\partial_z\ts=\epsilon_{ij}\partial_i\delta
n_j$ \refs{\NT,\MN}\
and can be absorbed into shifts in $K_A^{||}$, $K_2$ and $\bar C$.  Thus the
melted
crystal is precisely described by our liquid crystal free energy.

\newsec{The Uniform Conical Phase}
I
now look for ground states as a function of $K_A$, the amplitude of the hexatic
order
parameter.  The results of this section closely follow those found by Meyer for
a cholesteric in a magnetic field parallel to the pitch axis \MEYER .
The analysis here, however, involves the bond order parameter $\ts$ and its
equilibrium configuration as well as that
of the nematic director.  For simplicity I take $K_A^\perp=K_A^{||}=K_A$ and
likewise
take $\bar C=0$.  I consider as a class
of ground states the conical states proposed by Meyer \MEYER:
\eqn\evii{{\bf\hat n} =\left[\cos\phi\cos qz, \cos\phi\sin qz,
\sin\phi\right].}
where $q$ and $\phi$ are free constant parameters.
The equations of motion for $\ts$ are
\eqn\eviii{-K_A\nabla^2\ts - K_A\tilde q_0 \nabla\cdot{\bf\hat n} =0}
and so $\nabla^2\ts =0$ as ${\bf\hat n}$ is divergence-free.  Thus the only
solutions
for $\ts$ are linear functions of the co\"ordinates and so ${\bf
v}\equiv\nabla\ts$
is a constant
vector.  Inserting the {\sl ansatz} \evii\ and a constant vector ${\bf v}_0$
into \ei ,
\eii\ and \eiii , I have
\eqn\eix{F = \Omega\left\{{K_A\over 2}v_0^2 - K_A\tilde q_0
\sin\phi{\bf v}_0\cdot\hat z  +
{K_2\over 2}\left[q\cos^2\phi-q_0\right]^2 +
{K_3\over 2}q^2\sin^2\phi\cos^2\phi\right\}}
where $\Omega$ is the volume of space and the oscillating terms drop out upon
integration over space.  Minimizing with respect to ${\bf v}_0$ I find
\eqn\ex{{\bf v}_0\equiv\nabla\ts = \hat z\tilde q_0\sin\phi}
and thus, with $x\equiv\cos^2\phi$
\eqn\exi{F_{\rm eff} =\Omega\left\{{K_2\over 2}\left(qx - q_0\right)^2 +
{K_3\over 2}
q^2x(1-x) - {K_A\over 2}\tilde q_0^2
(1-x)\right\}}
This is precisely the energy studied by Meyer for a
cholesteric in a magnetic field of strength $H=\tilde q_0\sqrt{K_A}$.  In
\MEYER\ the
possibility of the cholesteric unwinding by changing $q$ was considered along
with the conical state.  In this case the former effect can never happen:
because ${\bf v}_0$
must be constant, the effective field will always be parallel to the pitch
axis.  Translating
Meyer's results I find that for $K_2<K_3$ there is a transition with increasing
$\sqrt{K_A}\tilde
q_0$ from the pure
cholesteric state ($\phi =0$, $q=q_0$) to a pure nematic state ($\phi=\pi/2$)
at
$\sqrt{K_A}\tilde q_0 = q_0\sqrt{K_2}$.  For $K_2>K_3$ there is an intermediate
phase as well.
For $\sqrt{K_A}\tilde q_0 < q_0\sqrt{K_3}$ the cholesteric state persists,
while
for $\sqrt{K_A}\tilde q_0 > q_0K_2/\sqrt{K_3}$ the nematic state persists.  In
between
\eqn\exii{q= \tilde q_0 \sqrt{K_A/K_3}}
and
\eqn\exiii{\cos^2\phi\equiv x = {K_2\sqrt{K_3/K_A}(q_0/\tilde q_0) - K_3\over
K_2-K_3}}
for $q_0\sqrt{K_3}\le \sqrt{K_A}\tilde q_0\le q_0K_2/\sqrt{K_3}$.  Thus the
conical phase
should appear continuously as a function of $\tilde q_0$, or, alternatively, as
a function of $K_A$ for fixed $\tilde q_0$.  Hence as the hexatic order grows
the cholesteric
state will become conical and, eventually, nematic.  The conical state has, in
addition
to a smectic-$C^*$-like director texture a rotating bond order as well, with
pitch
$\tilde q_0$.  If $\tilde q_0/q_0$ is not rational, the fluctuations around
this
ground state should be akin to those in an incommensurate smectic
\nref\RT{S.~Ramaswamy
and J.~Toner, J. Phys. suppl. {\bf A} (1990) 275.}\refs{\TONSM,\RT}.

It is unlikely that $\tilde q_0=0$ as it is allowed by the same symmetry that
allows the
cholesteric coupling $q_0$.  However, even if this is the case I expect
$\bar C\ne 0$.  In the case $\tilde q_0=0$, the $\bar C$ acts as a generated
$\tilde q_0$, in some sense.  I will still have a
magnetic field type term, though now it depends on
$q$ and $\phi$, which should produce a conical phase.  The phase diagram
at $\tilde q_0=0$ will depend on all of $K_2$, $K_3$, $K_A$ and $\bar C$, and
mapping it out would be straightforward yet tedious.

Turning back to $\tilde q_0\ne 0$,
we have found that the conical phase can persist for $K_2>K_3$ which is usually
not the experimental case.  However, in light of recent work on the fluctuation
enhancement of elastic constants \refs{\TONER,\KT}\ one might hope that this
inequality could always be met.  Indeed since there is no constraint
locking $2\partial_z\ts$ with $\nabla_\perpp\times\delta\vec n$ the non-chiral
part of the free energy from \KT\
would be
\eqn\exiv{F={1\over 2}\int d^d\!x\,\left\{E\left[\nabla_\perpp\cdot\vec u -
{1\over
2}(\partial_z\vec u)^2 +\omega\right]^2 + G\rho_0^2(\partial_z\omega)^2
+\free_{\hat
n}[\partial_z\vec u]\right\}}
where $E$ is the bulk compression modulus, $\rho_0$ the average polymer
density, $\partial_z\vec u
=\delta\vec n$, $G=kT\ell/\rho_0$, $\ell$ is the typical polymer length and
$\free_{{\bf\hat n}}$
is expanded to quadratic order in $\delta\vec n$ where ${\bf\hat n}\approx \hat
z+\delta\vec n$.
It was found in \KT\ that for polymers
longer than $\ell_0$, $E$, $K_2$ and $K_3$ took on anomalous,
$\ell$-dependent values.  For polymers with only steric entropic interactions
$\ell_0=L_P^3/a^2$
where $L_P$ is the polymer persistence length and $a$ is the average areal
interpolymer
spacing.  It was found that $K_2(\ell)\sim (\ell/\ell_0)^{0.20}$ and
$K_3(\ell)\sim(\ell/\ell_0)^{0.15}$ to two loops in a $d=4-\epsilon$ expansion
at
$d=3$.

To the free energy in \exiv\ I add the hexatic stiffness $\free_{\ts}$.
Expanding
around ${\bf\hat n} =\hat z$ I can consider the possible corrections to the
hexatic stiffness $K_A$ as well as the chiral couplings $q_0$ and $\tilde q_0$.
Following the analysis in \KT ,
I find that $K_A$ and $\tilde q_0$ do not
anomalously renormalize and suffer only finite shifts in their values.
To illustrate, I consider adding $\free_{\ts}$ to \exiv\ with $K_A^\perp=
K_A^{||}=K_A$.
I am interested in fluctuation-induced changes to $K_A$ and $\tilde q_0$.
As in the argument of \KT\ for the non-renormalization of $q_0$,
I consider the $\tilde q_0\approx 0$ limit.  Thus
I do not consider corrections to quadratic propagators arising from the
coupling $K_A\tilde q_0 \partial_z\vec u\cdot\nabla_\perpp\ts$.  It is
easy to see that, since I consider only corrections linear
in $\tilde q_0$, $K_A$ does not acquire any infinite renormalization, and,
in fact,
even if I consider $K_A^\perp\ne K_A^{||}$ the corrections from fluctuations
in $\dnb$ only lead to finite renormalizations.  The corrections
to $\tilde q_0$ must come from non-linearities in the chiral term itself;
no other term can generate the appropriate chiral coupling.  The tadpole
graph arising from this is proportional to $\langle\dnb^2({\bf r})\rangle$
and thus only contributes to a finite shift in $K_A\tilde q_0$ in three
dimensions, where nematic order persists.
Alternatively, by a judicious choice of rescalings \ref\TCL{I thank
Tom Lubensky for many discussions on this point.}, one can show
that the longitudinal part of $\vec u$ is irrelevant to the
renormalization of the Frank constants, and thus the new coupling,
which only involves the longitudinal part of $\vec u$, will not change
the
results in \KT .  Thus the mean field analysis of the uniform conical
state holds when I replace the constant Frank constants with
$\ell$-dependent Frank constants.
Hence, for sufficiently long polymers, it may be possible
to have the unusual situation of $K_2>K_3$, though since $K_2/K_3 \sim
(\ell/\ell_0) ^{0.05}$,
a twofold increase in the ratio of Frank constants implies the polymer must
be $10^6\ell_0$.  For instance, in DNA $L_P= 600\angstrom$ and $a=35\angstrom$
lead to $\ell_0 \approx
18 \mu m$.  With typical chain lengths on the order of centimeters, this can
lead to
a $37\%$ increase in the ratio of $K_2$ to $K_3$.  This increase leads one to
hope that the conical phase could persist and could be observed.

Aside from the light scattering behavior of this phase due to its N+6 character
\TON , it
should be possible to see this phase through crossed polarizers.
As hexatic order
grows past the cholesteric-to-conical point, the extinction directions
(perpendicular
to the director) will {\sl continuously} change from the nematic direction
to the (perpendicular) cholesteric plane.  This will happen without a density
modulation forming, as in
the smectic-$C^*$ phase which would have a similar behavior.  The density
modulation
could be detected (or not) via electron microscopy and x-ray scattering, thus
distinguishing
this new phase from the smectic.  As in the moir\'e phase of
chiral columnar crystals \KN, in
each constant $z$ cross section
the structure function will be, in Fourier space, that of a hexatic --
six broad spots.  When looking at the structure of the bulk sample, however,
the rotation of the bond order will merge these spots into a ring
in the $q_x$-$q_y$ plane.  The periodicity
of the rotation will lead to additional rings at $q_z=6n\tilde q_0\sin\phi$
where $n=\pm 1,\pm 2,\ldots$.

In addition, as in a smectic-$C^*$, this uniform conical texture
can exhibit ferroelectric behavior through a term in the free energy
\eqn\eflc{\free^*_{\rm flc} = \mu\left({\bf v\dot\hat n}\right){\bf E}\dot
\left({\bf v\cross\hat n}\right)= \mu\tilde q_0\sin\phi\cos\phi\left[E_y\cos
qz - E_x\sin qz\right]}
This extra interaction preserves the nematic symmetry and is chiral, since
under parity, $\bf E\rightarrow -E$.  I have taken the ``layer normal'' to
be $\bf v$ in constructing \eflc .  There is, however, a term that
is allowed in these chiral phases that has no analogue in the smectic-$C^*$.
Since the direction of $\bf v$ is determined by the chiral coupling $\tilde
q_0$
and $\bf\hat n$, a non-chiral term is allowed
\eqn\enflc{\free_{\rm flc} = \bar\mu{\bf E}\dot\left({\bf v\cross\hat n}\right)
}
which preserves the nematic symmetry ($\bf\hat n$ and $\bf v$ {\sl both}
change sign) and is not chiral.  Under spatial reflection $\bf v\cross\hat n$
unambiguously changes sign, as does $\bf E$.  This term is allowed
because there is a difference, for fixed $\bf\hat n$, between a
configuration with $\bf v$ and $-\bf v$.  They differ by the handedness
of the hexatic helix.  In a smectic-$C^*$ the layer normal does
not have an unambiguously defined direction because of the
up-down symmetry of the layered structure.
Unlike a smectic,
this ferroelectric liquid crystal does not have a density wave as in
the usual ferroelectric smectic liquid crystals.  The wave that supports the
conical texture is one
of twisting hexatic order along $\bf v$.  In section 5 I will make
the analogy between smectics and chiral hexatics more precise.  Note
that $\free_{\rm flc}$ also implies that an applied electric field
will favor configurations in which the hexatic order twists in
planes {\sl perpendicular} to $\bf\hat n$.

\newsec{Topological Constraints: The Mermin-Ho Relation and Saddle-Splay}

Up to this point I have ignored the topological relation between $\ts$ and
$\bf\hat n$.
{}From the definition
of $\ts$ arising from \eangle , it is clear that textures in $\bf\hat n$
can influence the local value of $\ts$.
The two orientational order parameters
are related to each other in much the same way that the two order
parameters
of $^3$He-A are related, namely, through the Mermin-Ho relation \MH .
I will
sketch here a brief derivation, following \VOL .
Let $\left\{{\bf\hat e}_1,{\bf\hat e}_2,
{{\bf\hat n}}\right\}$ be a right-handed, orthonormal triad.
If I consider a rotation $\delta\ts$ about ${\bf\hat n}$, then
\eqn\erota{
{\bf\hat e}_1' + i{\bf\hat e}_2' = e^{i\delta\ts}\left({\bf\hat e}_1 +
i{\bf\hat e}_2\right)
}
and so
\eqn\erotaii{
\eqalign{{\bf\hat e}_1' - {\bf\hat e_1} &\approx -\delta\ts{\bf\hat e}_2\cr
{\bf\hat e}_2'-{\bf\hat e}_2 &\approx \delta\ts{\bf\hat e}_1\cr}}
Thus $\delta\ts \approx {\bf\hat e}_1\cdot\delta{\bf\hat e}_2$, or
\eqn\erotaiii{
v_\mu
\equiv\nabb_\mu\ts = e_1^\alpha\nabb_\mu e_2^\alpha}
Taking the curl of both sides and using ${\bf\hat e}_1\cross{\bf\hat e}_2
={{\bf\hat n}}$,
we
have
\eqn\emh{[\nabb\cross {\bf v}]_\mu = {1\over
2}\epsilon_{\mu\nu\rho}\epsilon_{\alpha\beta\gamma}n^\alpha\partial_\nu
n^\beta\partial_\rho
n^\gamma,}
the celebrated Mermin-Ho relation.  The content of this relation is that in the
ground
state of the N+6 phase, disclinations in $\ts$ are forced in by the nematic
texture.
It quantifies a simple topological
consequence of non-cylindrically-symmetric molecules.  For concreteness, take a
biaxial molecule
in a cylindrical
geometry with normal boundary conditions. One might imagine
a $+1$ disclination line will run
up the center of the cylinder.  However, it is well known that such a geometry
will ``escape
into the third dimension'' along the disclination.  However, if the molecules
have
the same (biaxial) orientation at the cylinder wall, when they escape they will
be compelled to force a disclination into the biaxial order parameter in the
center.
This is illustrated in Figure 1.
This frustration is summed up in \emh .  Note that in the previous section, I
took
$\nabla\cross{\bf v} =0$, which, in the case of the Meyer {\sl ansatz}, is
justified
by simply plugging \evii\ into \emh .

When I look for ground state
configurations I must take the constraint into account.
The excited states differ from the ground state by the energy of a free
disclination.  In the hexatic phase, this energy diverges with the
system
size, so the Mermin-Ho relation should be enforced.  This is not unlike hexatic
phases
on flexible membranes.  There disclinations were forced in by gaussian
curvature \NELii .  In particular, if I take as the the hexatic director
${\bf\hat N} = \cos\ts{\bf\hat e}_1 + \sin\ts{\bf\hat e}_2$ and compute
the covariant derivatives of $\bf\hat N$ in the local basis determined by
$\left\{{\bf\hat e}_1,{\bf\hat e}_2, {\bf\hat n}\right\}$, I find
\eqn\ecov{\eqalign{
D_\mu N_1 &= -{\bf\hat e}_\mu\dot\left[\nabb\ts -
\bold{\Omega}\right]\sin\ts\cr
D_\mu N_2 &= {\bf\hat e}_\mu\dot\left[\nabb\ts -
\bold{\Omega}\right]\cos\ts\cr}}
where ${\bf\hat n}\equiv{\bf\hat e}_3$ and $\Omega_\mu=e_1^\nu\partial_\mu
e_2^\nu$.
The appropriate elastic free energy would be
\eqn\emhf{\free_{\Omega} = {k^\perp\over 2}\left[D_iN_j\right]^2 + {k^{||}\over
2}
\left[D_3N_j\right]^2 = {k^\perp\over 2}\left(\nabb\ts-\bold{\Omega}\right)^2
+{k^{||} - k^\perp\over 2}\left[{{\bf\hat n}}\dot\left(\nabb\ts -
\bold{\Omega}\right)\right]^2}
In terms of $\bf\hat N$, the term ${\bf\hat n}\dot\nabla\ts$, for instance,
becomes
$n_\mu\epsilon_{ij}N_iD_\mu N_j\equiv {\bf\hat
n}\dot\left(\nabla\ts-\bold{\Omega}\right)$.
Thus, in general, one should {\sl always
include} $\bold{\Omega}$ with derivatives of $\ts$ as a consequence of the
wobbling
about of ${\bf\hat n}$.  Because a change of basis vectors
${\bf\hat e}_1({\bf r})$ and ${\bf\hat e}_2({\bf r})$ should not
change the physics of $\ts$, I will consider only ``gauge-invariant''
correlations of $\ts$ through
\eqn\egits{G_{\ts\ts}({\bf r};\Gamma)=
\langle\,e^{6i\ts({\bf r})}\exp\{
i\int_\Gamma\bold{\Omega}\dot d\bold{\ell}\}
e^{-6i\ts({\bf 0})}\,\rangle}
where $\Gamma$ is a path connecting $\bf r$ and $\bf 0$.  When this
Greens function is non-vanishing for large separations $\bf r$, there
is long-range hexatic order.  This correlation
function will be invariant under shifts of $\bold{\Omega}$ by $\nabla\omega$
and $\ts$ by $\omega$,
where $\omega$ is an arbitrary function of space.  These transformations
are precisely those which do {\sl not} change $\curl\bold{\Omega}$ and
hence do not change the disclinations structure of $\ts$.  When $\curl
\bold{\Omega} = {\bf 0}$ everywhere, there are no obstructions to
choosing ${\bf\hat e}_i({\bf r})$
to be constant in space, leading to $\Omega={\bf 0}$ and to the usual
definition and interpretation of $\ts$.  Note that when the nematic is highly
disordered, with a rich wobbling texture, the contribution to \egits\ from
the nematic curvature can greatly reduce, and possibly destroy, long-range
hexatic order.

Since unbound disclinations cost a logarithmically-divergent
energy they will not appear in equilibrium.  However
regions where
$\nabb\times\bold{\Omega}\ne{\bf 0}$ have precisely the same
divergent energy as a free disclination.
Writing $\ts = \ts^{\rm smooth} +
\ts^{\rm sing}$
we can expand around a background of disclinations chosen to
screen these regions with $\nabla\ts^{\rm sing} =\bold{\Omega}$, leaving only
smooth
variations in $\ts$.  The dislocations are locked to $\nabb\times\bold{\Omega}$
according to the Mermin-Ho relation \emh .
Notice that in the case that ${{\bf\hat n}}$ corresponds to a field
of normal vectors to surfaces, ${{\bf\hat n}}({\bf
r})\dot\left[\nabb\cross\bold{\Omega}({\bf
r})\right]$ is
the Gaussian curvature of the surface at $\bf r$ to which ${{\bf\hat n}}({\bf
r})$ is
normal.  Rewriting the curvature, I find that ${{\bf\hat n}}
\dot\left(\nabb\cross\bold{\Omega}\right)\equiv -\nabla\dot\left[\left({\bf\hat
n}\dot
\nabla\right){\bf\hat n} - {\bf\hat n}\left(\nabla\dot{\bf\hat
n}\right)\right]$, the
well known saddle-splay term of nematic liquid crystals.  Thus I can identify
$\bar C'$ with $K_{24}$ the saddle-splay elastic constant.

This can be understood geometrically: to lowest order in $\dnb$
\eqn\mho{
[\nabb\cross\nabb\ts]_i= {1\over 2}\epsilon_{ijk}\epsilon_{\alpha\beta\gamma}
n^\alpha\dj n^\beta\dk n^\gamma\approx\epsilon_{ijk} \dj\dn_x\dk\dn_y}
Disclinations pointing along $\hat z$ are constrained by
\eqn\mhoz{
[\nabb\cross\nabb\ts]_z = \dx\dn_x\dy\dn_y - \dx\dn_y\dy\dn_x}
in other words, disclinations must
appear if the
directors are the normals to saddle-surface.  The saddle-splay
term measures the amount of saddle-like deformations in the director.  Since
$\ts$ is single-valued on a surface, the Mermin-Ho relation tells us that
the saddles must be such that the integral of their saddle-splay
around their boundary is a multiple of $2\pi/6$.  However, in the N+6
case, the director ${{\bf\hat n}}$ need not be the collection of normals to a
surface.  Indeed,
in the double twist texture of liquid crystal blue phases there is no surface
to
which the directors are normal.  If such a texture were present in the chiral
N+6 system, \mhoz\ implies that as a circuit is traversed around a central
molecule
of a double twist cylinder, there would be fewer than $z=6$ nearest neighbors
since the
saddle-splay is positive.
Considering the director texture around the center of a double twist core,
Figure 2
shows $19$ braided polymers with the center one being straight.  It is not
necessary
that these be polymers, just that the nematic director be tangent to the lines.
This texture has double
twist in the nematic field
as found
in
blue phases of cholesteric liquid crystals \refs{\SWM,\KLEMi}.
The integrated trajectories are also
known to be the projections of great circles of $S_3$, the surface of a
$4$-dimensional
sphere, onto flat $3$-dimensional space.  The curvature of $S_3$ implies that
the
co\"ordination number is
less than $z=6$ \refs{\KLEMi,\KLEM} which is consistent with \mhoz .
This is illustrated in Figure 3.  Here
the molecular centers (represented by dots) lie on a local triangular lattice,
as in the hexatic phase.  When
the polymers are tipped over, the distance between their centers is necessarily
increased.  This reduces the allowed packing density leading to co\"ordination
numbers less than $z=6$.

\newsec{The Equivalence of Smectics and Chiral Hexatics}
\subsec{Derivation of Chiral Hexatic Free Energy}
I have shown that for $K_2>K_3$ a uniform state with a smectic-$C^*$ texture
can persist for sufficiently strong hexatic ordering.  A different scenario
is also possible, leading to a defect state, analogous to the Abrikosov flux
lattice
in superconductors and {\sl isomorphic} to the twist grain boundary phase
of chiral smectic-$A$ \TGB . Consider first the Landau theory
for an isotropic hexatic in a background nematic field
\eqn\ehexf{\free_{\rm hex}
= {1\over 2}\vert{\bf D}\psi_6\vert^2
+ r\vert\psi_6\vert^2 + u\vert\psi_6\vert^4}
where $\psi_6$ is the hexatic order parameter, which, in the broken phase, is
approximately
$\psi_6\approx(\sqrt{K_A}/6)e^{6i\ts}$, and,  as usual, $r\propto (T-T_c)$.
$\bf D$ refers to the
covariant derivative, which subtracts $\bold{\Omega}$ from the gradient of
$\ts$.

Adding to this the chiral term (covariantly modified)
in $\free_{\ts}$, I have
\eqn\ehexcf{\free_{\rm hex}^* =
6i{\tilde q_0}{\bf\hat n}\cdot
\left[\psi_6^*{\bf D}\psi_6 - \psi_6{\bf D}\psi_6^*\right] +
{36\over 2}\tilde q_0^2\vert\psi_6\vert^2\left({\bf\hat n}^2 -1\right)}
Recall that under ${\bf\hat n}\rightarrow -{\bf\hat n}$, $\psi_6\leftrightarrow
\psi_6^*$ and so the first term in \ehexcf\ respects the nematic symmetry.
The second term, though identically zero, will be split up so that I may write
the total free energy as
\eqn\etotf{F_{\rm hex} = \int d^3\!x\,\left\{{1\over 2}
\left\vert\left({\bf D} - 6i\tilde q_0
{\bf\hat n}\right)\psi_6\right\vert^2  + \left[r-18\tilde
q_0^2\right]\vert\psi_6\vert^2 +
u\vert\psi_6\vert^4 + \free_{\bf\hat n}\right\}}
Thus, after shifting $T_c$ upwards by $18\tilde q_0^2$,
the free energy of a chiral hexatic is precisely that of a chiral
smectic \ref\DG{P.G.~de~Gennes, Solid State Commun. {\bf 14} (1973) 997.}
where $2\pi/6\tilde q_0$, the spacing between surfaces with the same bond-order
orientation, is the equilibrium smectic layer spacing.
There is an important difference between the smectic and the chiral hexatic
which is what allowed the non-chiral ferroelectric
term \enflc\ in the latter.  In a
smectic, the physics is invariant under $\tilde q_0\rightarrow -\tilde q_0$
{\sl and} $\ts\rightarrow -\ts$ ($\psi_6\leftrightarrow \psi_6^*$).  While
that appears to be a symmetry of \etotf\ recall that $\ts\rightarrow -\ts$
{\sl must} be accompanied by ${\bf\hat n}\rightarrow -{\bf\hat n}$.  In other
words, there is a difference between $\tilde q_0$ and $-\tilde q_0$ in
the hexatic that does not exist in the analogous smectic.
It is also important to note the distinction between
$\bold{\Omega}$ and $\bf\hat n$.  Shifts of $\ts$ by $\omega$ accompanied by
shifts of $\bold{\Omega}$ by $\partial\omega$ are true gauge transformations.
The
invariance under these transformations is not the sign of a symmetry, but
rather
a redundancy of identical physical descriptions.  Rigid rotations in space, on
the
other hand, are
responsible for the gauge-{\sl like} coupling of $\ts$ to $\bf\hat n$.  Under
rotation by
angle $\bold{\Phi}$, ${\bf\hat n}\rightarrow {\bf\hat n} +
\bold{\Phi}\cross{\bf\hat n}$
and $\ts\rightarrow \ts + \left[\bold{\Phi}\cross{\bf\hat n}\right]\cdot{\bf
r}$, so
that, to lowest order in fluctuations of $\bf\hat n$ around a
spatially-uniform,
ordered state ({\sl e.g.}, ${\bf\hat n} = \hat z$), the rotation is
a symmetry of the theory.  A rotation is physical and corresponds to different
but equivalent physical systems \ref\NH{T.C.~Halsey and D.R.~Nelson, Phys. Rev.
A {\bf 26}
(1982) 2840.}.  In the highly aligned limit, when $\dnb$ is small and
$n_z\approx 1$,
$\curl\bold{\Omega} = \bo{\dnb^2}$ and hence the contribution from
the covariant derivative
is much smaller than that from the chiral term in the
derivative $\tilde q_0\dnb$.  Thus I expect that the energetics
of the chiral hexatic phases should be similar to that of the analogous
smectic phases.

It is now clear that the uniform conical state is just one possible smectic
phase.  In the nematic state ($\sin\phi=1$), there is pure chiral hexatic
order, with $\ts=\tilde q_0 z$.  This is equivalent to a smectic-$A$ phase
as shown in Figure 4.
Upon reduction of the nematic order, the nematic director tips out of
the $\hat z$ direction and the phase is equivalent to a smectic-$C^*$
phase.  The chiral order acquires a longer pitch, which is equivalent
to a larger smectic layer spacing.
Finally, as the hexatic order is reduced further, the director
lies down into a nematic texture and there is no chiral hexatic order.  This
is just a pure cholesteric.  The smectic layer spacing has gone to infinity
and there is no smectic order.

\subsec{Defect Phase: The Renn-Lubensky State}
Ignoring the complications of the Mermin-Ho relation and the covariant
derivative,
this model has been studied extensively \TGB\ and, in the type-II regime (where
$u$ is
sufficiently large) one expects a ground state with a proliferation of defects.
 I propose
then a new twist grain boundary state of {\sl hexatics}.  It will consist of
regions of
twisted N+6 separated by grain boundaries made up of hexatic disclinations.
Within
in each region the bond order will twist with pitch $2\pi/\tilde q_0$ and
the director will be well aligned in a nematic state. As one moves
along a pitch axis perpendicular to the nematic director, across a grain
boundary,
the nematic direction will jump by some finite angle $\alpha$.  This will lead
to a state which, at long distances, will appear to be pure cholesteric, but
will in
fact have regions with rotating hexatic order.

Since the possibility of the defect phase is independent of the existence
of the uniform conical phase, it is possible that the uniform conical phase
could be punctuated itself by defects.  In this case $\phi_\infty$ would be
the equilibrium conical angle and the director would relax to this cone.  This
exotic
phase would be similar to the ${\rm TGB}_C$ phase \ref\RL{S.R.~Renn and
T.C.~Lubensky,
Mol. Cryst. Liq. Cryst. {\bf 209} (1991) 349.} in which the defect-free regions
are not smectic-A, but rather smectic-C, though in this case it may be more
appropriate
to think of the clean regions as being smectic-C$^*$ instead.

One might speculate on other possible phases.  Since the bond order field $\ts$
has much the same behavior as $\theta_2$, a biaxial order parameter, structures
of chiral N+6 phases should be in close analogy with the myriad of blue
phase textures:  in other words an N+6 phase is closely related to a biaxial
nematic.  Owing to the connection with smectics it is a savory proposition to
think of blue phases made of smectic layers, where now the phase $\theta_2$
present in the high-chirality limit of
a blue phase \SWM\ could be interpreted as ticking off the lamellar layers of
the
smectic in space.  This prospect is under investigation \ref\KLBP{R.D.~Kamien
and T.C.~Lubensky, unpublished.}.

\subsec{The Structure of Topological Defects}
In light of the Mermin-Ho relation, the structure of the defects requires some
clarification.
Configurations which satisfy the Mermin-Ho relation can have disclinations in
$\ts$
without a logarithmically divergent energy per unit length.  However,
configurations
are not required to be so innocuous.  The defects necessary for a TGB state are
still possible.  The energy of the core will be modified, of course, by the
extra energy associated with the ``curvature'' of the nematic field.
Considering a screw dislocation in the hexatic order, $\ts$ will wind around in
each
$xy$-plane, independent of $z$, and so $\ts = \theta$, where $\theta$
is the azimuthal co\"ordinate.  The equations of motion imply
that at radii larger than $\sqrt{K_1/K_A}/\tilde q_0$ the
nematic field will lock to the ``superfluid velocity'' ${\vec
v}=\nabla_\perpp\ts/\tilde q_0$, and
thus
\eqn\ednoc{\dnb \approx {1\over \tilde q_0 r}\hat\theta}
where $\hat\theta\equiv (x\hat y -y\hat x)/r$ is the unit azimuthal
vector in cylindrical co\"ordinates.  Note that since $\dnb$ does
not depend on $z$ the only component of the Mermin-Ho relation that
is non-zero is the $z$ component.  Using \mhoz\ I find, to
leading order in $\dnb$, that
\eqn\etheththt{\left[\curl\bold{\Omega}\right]_z = -{1\over \tilde q_0^2}
{1\over r^4}}
Thus for large $r$ the gauge field $\bold{\Omega}$ falls off
like $r^{-3}$, which is much faster than the rate at which the
vorticity falls off ($r^{-1}$) and equal to the rate at
which the difference $(\nabla_\perpp\ts - \tilde q_0\dnb)$ falls off at
infinity.  Thus, ignoring the effect of the nematic curvature adds an
energy comparable to the energy of the defect itself, and certainly does
not cause the defect to become energetically prohibited.

Turning back to the Meyer {\sl ansatz}, one might consider defects
in the uniform conical state.  I can solve
the Mermin-Ho relation to find a family of defects in a conical state.
Using a radially dependent
{\sl ansatz}, I take
\eqn\espa{{\bf\hat n} = \left[\cos\phi(r)\cos q_0z,\cos\phi(r)\sin
q_0z,\sin\phi(r)\right]}
In the center of the defect, $\phi(0)=0$ and I have a pure cholesteric region
--
a region which does not support hexatic order.  Moreover, at infinity, the
director relaxes
to a conical state, {\sl i.e.} $\phi(\infty)=\phi_\infty$.  In this case, it is
straightforward
to compute
\eqn\ecurls{\eqalign{\left[\curl{\bf v}\right]_x &= -{yq_0\cos\phi\over
r}{\partial \phi\over
\partial r}\cr
\left[\curl{\bf v}\right]_y &= {xq_0\cos\phi\over r}{\partial\phi\over \partial
r}\cr
\left[\curl{\bf v}\right]_z &= {\cos\phi\over r} {\partial\phi\over \partial
r}\cr}}
solving for ${\bf v}$ I have
\eqn\esolvfv{{\bf v} = {\sin\phi(r)\over r}\hat\theta +
+q_0\left(\sin\phi_\infty-\sin\phi(r)\right)
\hat z + \nabla\ts}
where $\hat\theta\equiv (x\hat y - y\hat x)/r$
is the unit azimuthal vector in cylindrical co\"ordinates, and $\nabla\ts$
is a non-singular variation of the bond-angle direction, with equilibrium
value $\tilde q_0\sin\phi_\infty\hat z$ in a pure conical state.
Far from the core, as $r\rightarrow\infty$, ${\bf v}\rightarrow\nabla\ts=
\tilde q_0sin\phi_\infty\hat z$
and is
not singular.  Inside the core, $\sin\phi\approx 0$ and, minimizing
the free energy for $\nabla\ts$, the bond order is constant.
To be more precise,
in order to have nonsingular field configurations, $\curl{\bf v}$ must not
diverge
at the origin.  Since $\cos\phi(r\rightarrow 0)=1$, \ecurls\ implies that
$\partial_r\phi$ must go to $0$ no slower than linearly in $r$.  Hence
$\sin\phi(r)/r$
is well behaved at $r=0$.  In the core of this defect there is pure cholesteric
order of the nematic field, while outside the core there is a pure
uniform conical state.

In considering configurations \esolvfv ,
it is essential to note that, when integrated around a closed loop of radius
$R$
in an arbitrary constant-$z$ slice, $\ts$ must be single-valued up
to shifts by $2\pi/6$ and hence
\eqn\ethg{{2\pi n\over 6}=\int_{\partial M} d\vec\ell\dot{\bf v}=
\int_M dxdy\,\left[\curl{\bf v}\right]_z = 2\pi\left[\sin\phi(R)
- \sin\phi(0)\right]}
Thus, if I consider configurations without any free disclinations, then
$\sin\phi(r)$
must take on values which are integer multiples of $1/6$.  This would prevent
$\bf\hat n$ from relaxing back to a nematic configuration on a longer length
scale
than the disappearance of $\vert\psi_6\vert$.  In other words, the quantization
of
the circulation of $\ts$ enforces a quantization in the tilt of $\bf\hat n$.
Therefore,
if there were a core in which $\psi_6$ vanished (of radius $\sim\xi$), outside
that
core the nematic would have to have a constant tilt. This is as if
the London penetration
depth $\lambda$ were smaller than $\xi$.
This prevents the usual type II defects of the TGB state.
This sort of restriction is reminiscent of the disclination buckling transition
in
hexatic membranes
\ref\SE{S.~Seung and D.R.~Nelson,
Phys. Rev. A {\bf 38} (1989) 1005; S.~Seung, Ph.D. Thesis, Harvard University
(1990); see
also \NELii.}.  This type of defect has a core in which
there is not hexatic order, or perhaps an annular region in which the tilt
of the nematic changes gradually by $2\pi m/6$, for integer $m$.  Outside the
core there is hexatic order as well as nematic order. Inside there is a gain in
cholesteric energy as the nematic tips over and exploits its chirality, while
there is a loss of hexatic energy as its order is destroyed.  A detailed
energetic
calculation, using the entire rotationally invariant theory must be performed
in order to determine whether this non topological defect is stable.

\newsec{Acknowledgments}
It is a pleasure to acknowledge stimulating discussions with P.~Aspinwall,
M.D.~Goulian,
T.C.~Lubensky,
D.R.~Nelson, P.~Nelson, and J.~Toner.
RDK was supported in part by the National Science Foundation, through Grant
No.~PHY92--45317.

\footatend\vfill\supereject\immediate\closeout\rfile\writestoppt
\baselineskip=14pt\centerline{{\bf References}}\bigskip{\frenchspacing%
\parindent=20pt\escapechar=` \input refs.tmp\vfill\eject}\nonfrenchspacing
\nfig\fone{Escape into the third dimension of a biaxial object.  As a $+1$
nematic
disclination tries to escape into the third dimension, the small axis
director is forced to have a $+1$ disclination.  The long thicker line is
the nematic axis while the shorter, thinner line is the small director.
At the boundary the nematic is normal and the small director is uniform
and normal to the plane of the boundary.}
\nfig\fseven{A braided state with a double twist texture.  The center
line is straight and has fewer than $z=6$ nearest neighbors.  The lines
may represent long polymers or merely curves with tangent vectors
equal to the local nematic director (the integral curves).}
\nfig\feight{Illustration of the frustrated packing of tilted molecules.
Note that when the molecules
are tipped over, the distance between their centers increases.  This reduces
the
allowed packing density and leads to co\"ordination numbers less than $z=6$.}
\nfig\fff{Model of a chiral hexatic.  In each plane the bond-order
parameter is $\ts=\ts^0\;{\rm mod}\;2\pi/6$.  Between the planes the bond
order uniformly precesses along the average nematic director, ${\bf\hat n}
=\hat z$.  The planes are analogous to smectic planes, though there is no
density wave in this liquid crystalline phase.}

\vfill\eject\immediate\closeout\ffile{\parindent40pt
\baselineskip14pt\centerline{{\bf Figure Captions}}\nobreak\medskip
\escapechar=` \input figs.tmp\vfill\eject}

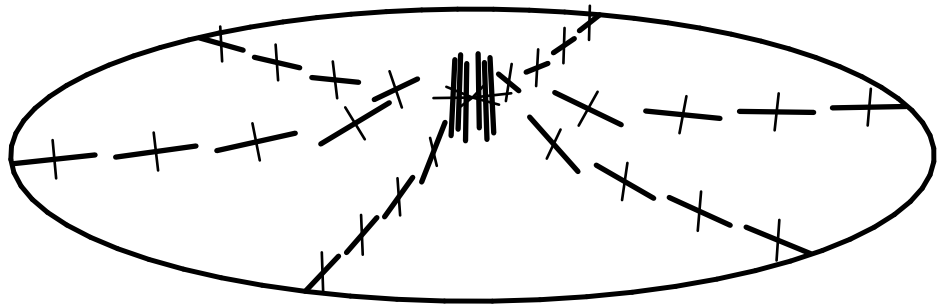
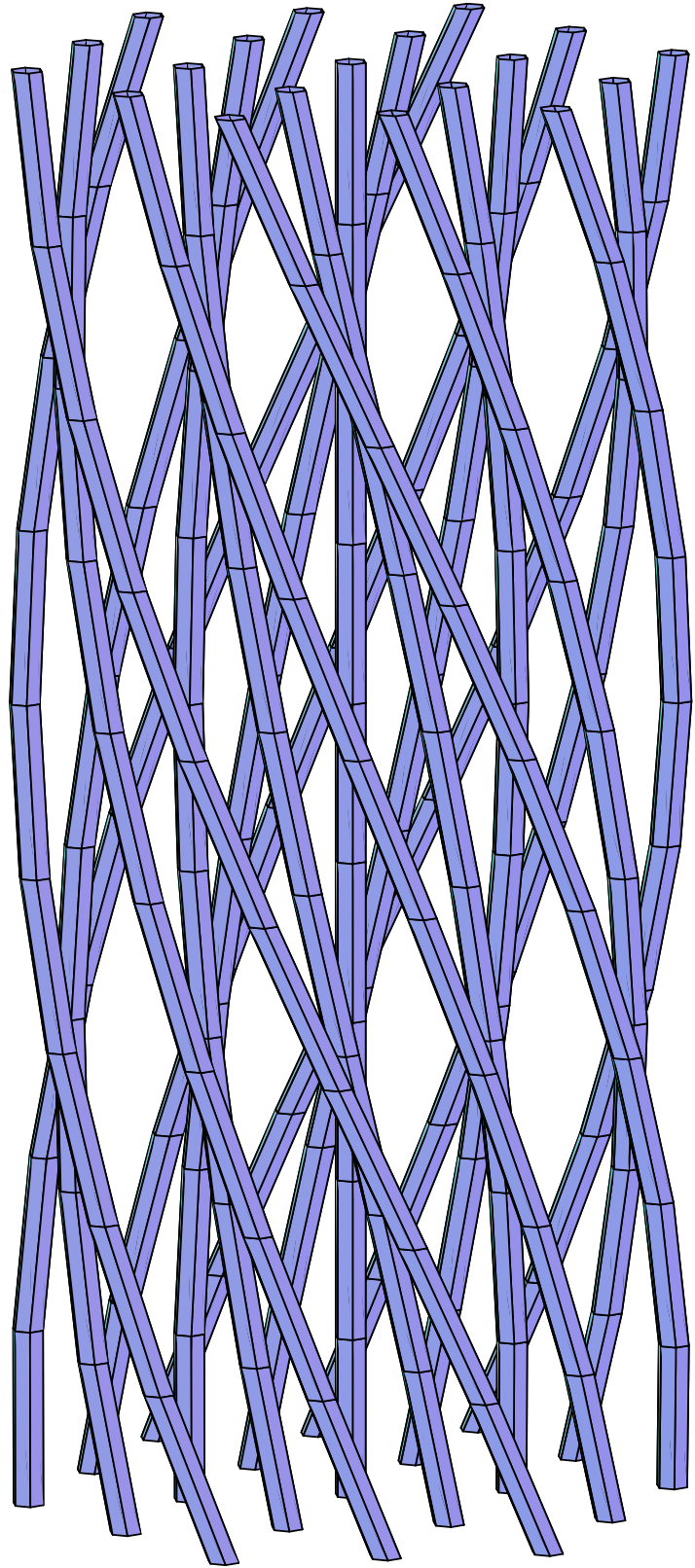
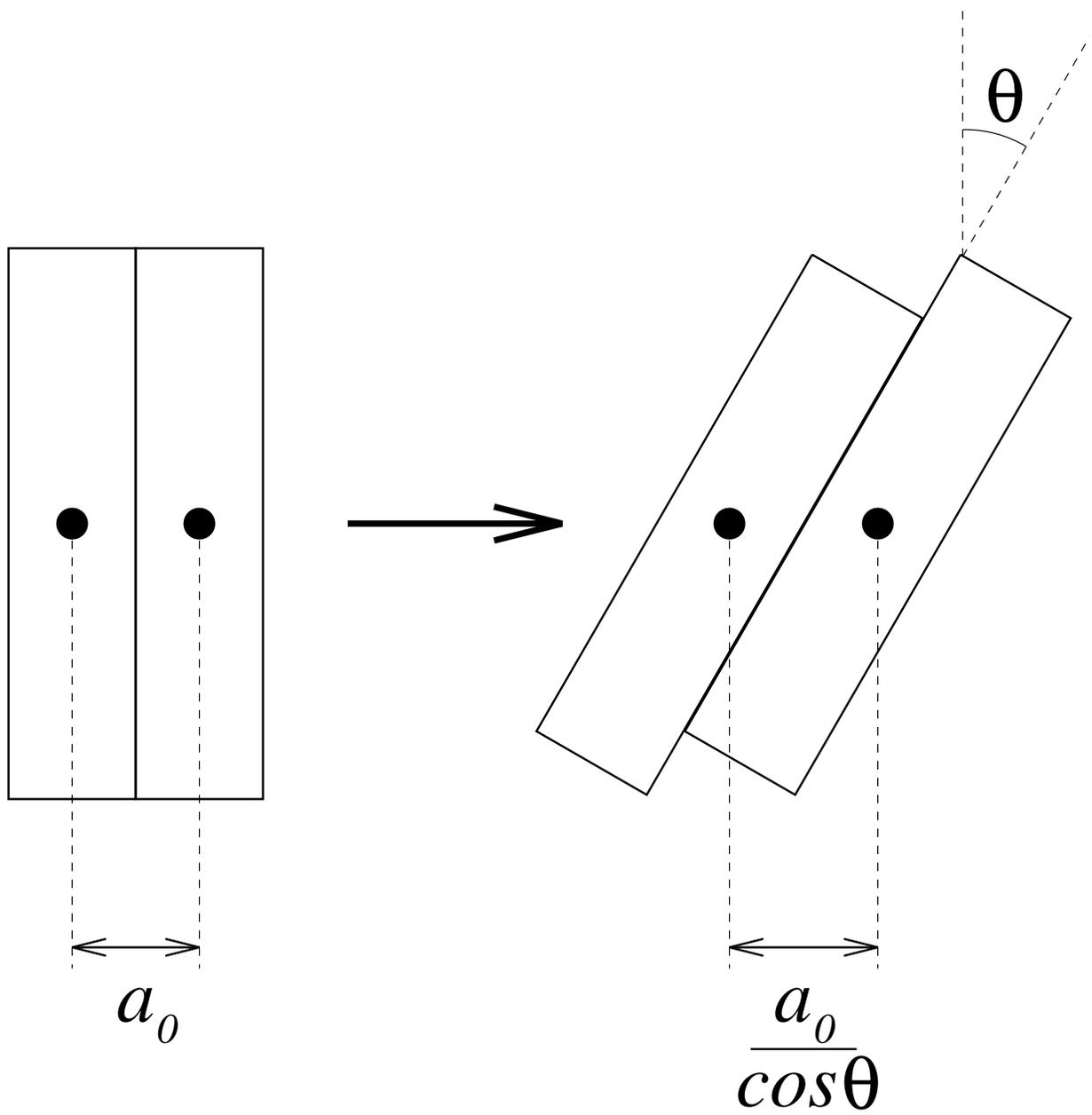
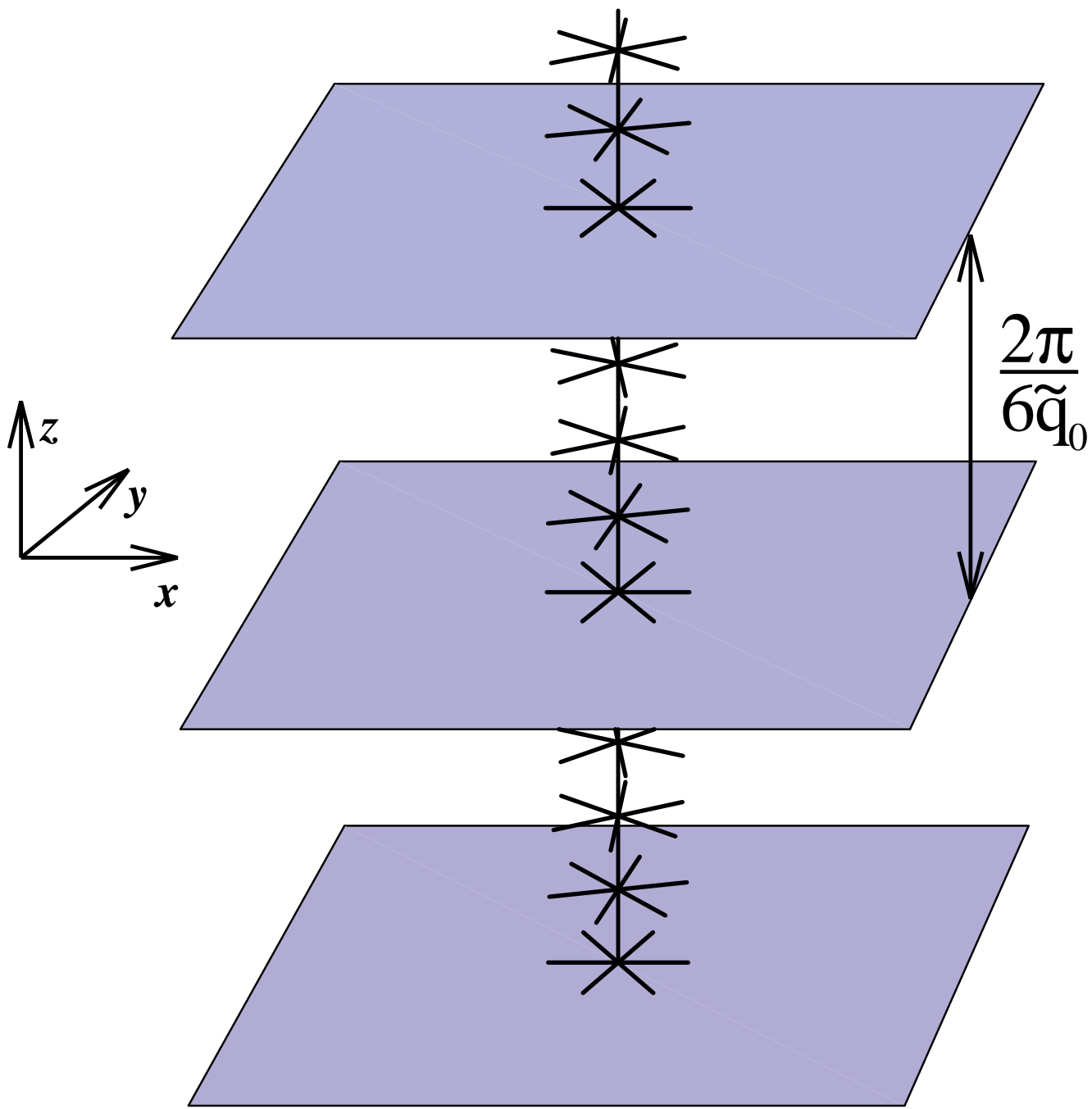
\bye